\documentclass[12pt]{article}
\usepackage{epsfig}
\usepackage{graphicx}
\usepackage{pifont}
\usepackage{txfonts}
\usepackage{hyperref}
\def \B{\bar{B}}
\def \branch{{\mathcal B}\,}

\newcommand{\tanbeta}{$\tan \beta$ }
\newcommand{\MSUSY}{$M_{\mathrm{SUSY}}$ }

\newcommand{\arXivid}[1]{{\href{http://arxiv.org/abs/#1}{arXiv:#1}}}

\newcommand{\cC}{{\cal C}}

\newcommand{\cB}{{\cal B}}
\newcommand{\wz}{\sqrt{2}}

\newcommand{\bit}{\begin{itemize}}
\newcommand{\eit}{\end{itemize}}
\newcommand{\mbo}[1]{$ #1 $ }
\newcommand{\tv}{\mbox{TeV}}
\newcommand{\gv}{\mbox{GeV}}

\newcommand{\power}[1]{\cdot 10^{#1}}
\newcommand{\epo}{\;.}

\newcommand{\cs}{\;,\;\;}

\newcommand{\Repa}{\mbox{Re} \:}
\newcommand{\amu}{a_\mu}
\newcommand{\be}{\begin{equation}}
\newcommand{\ee}{\end{equation}}
\newcommand{\ba}{\begin{eqnarray}}
\newcommand{\ea}{\end{eqnarray}}

\newcommand{\gapprox}{\gtrsim}

\newcommand{\braket}[1]{\langle{#1}\rangle}

\DeclareMathSymbol{\varPhi}{\mathalpha}{operators}{"08}
\DeclareMathSymbol{\varPsi}{\mathalpha}{operators}{"09}
\DeclareMathSymbol{\varOmega}{\mathalpha}{operators}{"0A}

\def\Title#1{\begin{center} { { #1} } \end{center}}
\usepackage{fancyhdr}
\begin{document}
\thispagestyle{empty}
\begin{flushright}
HU-EP-12/08 \\
DESY 12--037 \\
March 2012\\
\end{flushright}

\setcounter{page}{0}

\mbox{}
\vspace*{\fill}
\begin{center}
{\Large\bf Implications of low and high energy measurements on SUSY
models} \\

\vspace{5em}
\large
F. Jegerlehner\footnote[4]{\noindent 
Presented at Linear Collider 2011: 
Understanding QCD at Linear Colliders  in searching for old and new physics, 12-16 September 2011, ECT*, Trento, Italy.
This work
was supported in part by the European Commission's TARI program under contract
RII3-CT-2004-506078 and by the Organizers.} \\
\vspace{5em}
\normalsize
{\it Humboldt-Universit\"at zu Berlin, Institut f\"ur Physik,
       Newtonstrasse 15, D-12489 Berlin, Germany}\\
{\it Deutsches Elektronen-Synchrotron (DESY), 
Platanenallee 6, D-15738 Zeuthen, Germany}
\end{center}
\vspace*{\fill}
\newpage

\topskip 2cm

\Title{\bf \Large Implications of low and high energy measurements on SUSY
models}
\bigskip

\begin{raggedright}  

{\it \underline{Fred Jegerlehner}  \index{}
\footnote{Presented at Linear Collider 2011: 
Understanding QCD at Linear Colliders  in searching for old and new physics, 12-16 September 2011, ECT*, Trento, Italy}\\
Institut f\"ur Physik, Humboldt Universit\"at zu Berlin, Newtonstrasse 15, D-12489 Berlin, Germany\\
Deutsches Elektronen-Synchrotron (DESY), Platanenallee 6, D-15738 Zeuthen, Germany\\
{\rm  fjeger@physik.hu-berlin.de}
}

\bigskip\bigskip
\end{raggedright}
\vskip 0.5  cm
\begin{raggedright}
{Abstract}\\ New Physics searches at the LHC have increased
significantly lower bounds on unknown particle masses. This increases
quite dramatically the tension in the interpretation of the data: low
energy precision data which are predicted accurately by the SM (LEP
observables like $M_W$ or loop induced rare processes like $B \to X_s
\gamma$ or $B_s \to \mu^+\mu^-$) and quantities exhibiting an observed 
discrepancy between SM theory and experiment, most significantly found
for the muon $g-2$ seem to be in conflict now. $(g-2)_\mu$ appears to be
the most precisely understood observable which at the same time
reveals a 3-4 $\sigma$ deviation between theory and experiment and
thus requires a significant new physics contribution. The hints for a
Higgs of mass about 125 GeV~\cite{ATLAS,CMS}, which is precisely what
SUSY extensions of the SM predict, seem to provide a strong indication
for SUSY. At the same time it brings into serious trouble the
interpretation of the $(g-2)_\mu$ deviation as a SUSY contribution.
\end{raggedright}

\section{Minimal Super Symmetric extensions of SM}
The Standard Model (SM), although doing surprisingly well in
describing most of the precision data at the quantum level, is incomplete
as it does not incorporate dark matter (DM) for example and it has
fine tuning problems most notably it predicts a vacuum energy
contribution induced by the Higgs condensate which is 50 orders of
magnitude too large\footnote{With the Higgs vacuum expectation value  $\varv =246.22~\gv$ and
a Higgs mass of about 125 GeV the contribution from
the Higgs mechanism to the vacuum density is $\rho^{\rm
vac}_H=\frac18\,m_H^2\,
\varv^2\simeq 1.1841\power{8}~\gv^4$~\cite{Dreitlein:1974sa}. With $\kappa=8\pi
G_N/c^2$ the contribution to the cosmological constant is given by
$\Lambda_H=\kappa\,\rho^{\rm vac}_H\simeq 5.1095
\power{-2}~\mathrm{cm}^{-2}$. This compares
to the observed value $\Lambda_{\rm obs}=\kappa\,\rho_{\rm crit}\,
\Omega_{\Lambda}
 \simeq 1.6517 \power{-56}~\mathrm{cm}^{-2}\epo$} and also the 
Higgs mass is not protected from being much heavier than other SM
particles. As we know all SM states are protected either by chiral or
by gauge symmetries, except from the Higgs. Supersymmetry (SUSY)
imposing an invariance under the exchange of bosons/fermions with fermions/bosons in a
field theory in principle is able to cure these problems. Exact SUSY
would not only predict a vanishing cosmological constant, as
$\braket{H}=0$ for a supersymmetric Hamiltonian~\cite{Haag:1974qh},
but also vanishing anomalous magnetic moments like
\mbo{a_{\mu\,\mathrm{exact SUSY}}=0}, or
\mbo{ (\bar{B} \to X_s \gamma)_{\mbox{\raisebox{0.3ex}
{\tiny exact SUSY}}}=0}~\cite{FerraraRemiddi74}.
Since the SM predicts positive values for these observables the SUSY
complement of the supersymmetric extension of the SM should yield
negatively interfering contributions of the same size. 

If supersymmetry is imposed, scalars have fermionic partners protected
by chiral symmetry and therefore also scalars are required to be
massless in the symmetric phase\footnote{In supersymmetric quantum
field theories if not all then at least the leading ultraviolet
singularities cancel. This stabilizes the relation between bare and
renormalized quantities. In particular the only quadratic divergences
exhibited in the SM, the Higgs mass renormalization, is then absent in
the symmetry limit. Note that conformal symmetry also could provide a
solution to the Higgs hierarchy problem. The argumentation refers to a
scenario where the renormalized theory is the long distance tail of an
underlying theory at short distances which is exhibiting a physical
cut-off.}. Not only the SM gauge symmetry is broken. As we know from
observation, any SUSY extensions of the SM must be broken in such a
way that all sparticles are heavier than all SM particles. Still, for
the Higgs a minimal SUSY extensions of the SM \underline{predicts} a
light Higgs \mbo{m_h < M_Z+
\mathrm{radiative \ corrections}\leq 140~\gv} and finding a Higgs
in this range is a strong argument in favor of SUSY (see the blue-band
plot Fig.~3 in~\cite{Buchmueller:2007zk}).

In broken SUSY scenarios, patterns present at the exact symmetry level often
are completely spoiled and radiative correction effects of either sign
and much enhanced relative to the SM are possible. Order of magnitude
enhancements of radiatively suppressed SM results, possible in $B_s
\to \mu^+\mu^-$ decay, for example, are the most
interesting possibilities. However, precision data largely constrain
the size of SUSY contributions as long as the SM predictions match the
data. A particular role here plays the muon $g-2$ because a 3 to 4
$\sigma$ contribution is substantial. One also could understand the
supersymmetrized SM as \underline{the} particular extension of the SM
which is able to hide the rich structure it predicts from producing
substantial observable effect below the 1 TeV scale.

A viable Minimal Supersymmetric extensions of the SM (MSSM) is
possible only if we supplement the SM with an additional Higgs doublet
(2HDM). One reason is supersymmetry itself, the other is anomaly
cancellation of the SUSY partners of the Higgses. We thus have the SM
with two scalars, a lighter $h$ and a heavier $H$, a pseudoscalar $A$
and the charged Higgses $H^{\pm}$ the spectrum of which is doubled by
the SUSY completion, the sparticles. The vacuum expectation values of
the two scalars $\varv_i=\braket{H_i}\,(i=1,2)$ define the new parameter
$\tan \beta =\varv_1/\varv_2$. In the minimal SUSY models the masses
of the extra Higgses at tree level are severely constrained by mass-
and coupling-relations. Only two independent parameters are left,
which we may choose to be $\tan \beta$ and $m_A$. Very important is
the fact that the SM gauge structure is not touched when going to the
MSSM and gauge and Yukawa couplings of the sparticles are completely
fixed by the gauge couplings of the SM.

In general 2HDMs do not exhibit the phenomenologically well
established ``minimal flavor violation'' (MFV) constraint, which
demands FCNC and CP patterns to be close to what we have in the
SM~\cite{MFV02}. The trick which saves the peculiar SM features is
R-parity, a $Z_2$ symmetry between the two Higgs doublet fields $H_1
\leftrightarrow H_2$. As a byproduct SUSY+R-parity implies a stable
lightest SUSY particle (LSP) which is a good candidate for the
astrophysically established dark matter. The LSP usually is the
lightest neutralino
\mbo{\tilde{\chi}^0_1}, but also a gravitino  can be
a viable DM candidate. At the LHC the existence of a LSP would be
signaled by events with missing transversal energy.

Even with the constraints mentioned, SUSY extensions of the SM allow for
a large number $\sim 100$ of free symmetry breaking parameters. 
Free parameters typically are masses and mixings of the neutralinos,
the higgsino mass \mbo{\mu}
 (\mbo{+\mu H_1H_2} term of the 2HDM Higgs potential)
and \mbo{\tan \beta}.

This changes if one marries SUSY with GUT ideas, in fact SUSY-GUTs
(e.g. as based on SU(5)) are the only theories which allow for grand unification
broken at a low scale ($\sim$ 1 TeV). This provides strong
constraints on the SUSY breaking mechanism, specifically we distinguish the
constrained CMSSM a SUSY-GUT with soft breaking masses universal at the
GUT scale. The NUHM is as CMSSM with non-universal Higgs masses.
Minimal super gravity (mSUGRA) exhibits super gravity induced SUSY
breaking with \mbo{m_{3/2}=m_0} at the bare level. These models assume many
degeneracies of masses and couplings in order to restrict the number
of parameters. Typically, SM parameters are supplemented by $ m_{1/2}$
(scalar-matter mass, like $m_{\tilde{q}}$, $m_{\tilde{\ell}}$),
$m_{0}$ (the $U(1)_Y\otimes SU(2)_L$
gaugino masses, $m_{\tilde{\gamma}}$, $m_{\tilde{Z}}$,
$m_{\tilde{W}}$ and gluino mass $m_{\tilde{g}}$), $\mathrm{sign}(\mu), \tan
\beta,A$ (trilinear soft breaking term), and more for less constraint models.

\section{Low energy monitor: the muon anomaly}
Formally $a_\mu$ is one of the simplest observables one can imagine,
just the electromagnetic vertex
{ $(-ie)\:\bar{u}(p')\left[\gamma^\mu
F_1(q^2)+i\frac{\sigma^{\mu\nu}q_\nu}{2m_\mu}F_2(q^2) \right]u(p)$}
in the static limit where \mbo{F_1(0)=1,\;\, F_2(0)=a_\mu}. And it
has a simple experimental consequence, it is responsible for the
Larmor precession of a muon circulating in a homogeneous magnetic
field and which can be measured very precisely.
Presently we have~\cite{newBNL,Jegerlehner:2009ry,Davier:2010nc,Jegerlehner:2011ti,g-2}
\be
a_\mu^\mathrm{Exp.} =1.165 920 80(63) \times
10^{-3}
~~~~~~a_\mu^\mathrm{The.} =1.165 917 97(60)
\times 10^{-3} 
\ee
\be
\delta a_\mu=a_\mu^\mathrm{Exp.}-a_\mu^\mathrm{The.}=(283\pm 87)\: \power{-11}\cs
\label{deviation}
\ee
which is a $3.3\:\sigma$ deviation. If we take quoted errors and
uncertainties to be estimated correctly and if we assume it is not a
statistical fluctuation\footnote{The statistical error of
$a_\mu^\mathrm{Exp.}$ is $54 \power{-11}$, other errors are
systematic.} we have to conclude that we see physics beyond the SM:
$\delta a_\mu=\Delta a^{\mathrm{NP}}_\mu$.

Nevertheless, the status of the theory must be examined. In particular
the estimates of hadronic effects is by no means always 100 \%
certain.  Recently, it has been shown that taking into account
properly $\rho-\gamma$ mixing, which is absent in $\tau$-decay,
actually accounts for the $\tau$ versus $e^+e^-$
discrepancy~\cite{Davier:2010nc}. It means that $\tau$ data have to be
corrected according to $\varv_0(s)=r_{\rho\gamma}(s)\,R_{\rm
IB}(s)\,\varv_-(s)$ with a mixing correction $r_{\rho\gamma}(s)$ which
had not been taken into account in previous
analyses~\cite{Jegerlehner:2011ti}. These findings have been
obtained/confirmed in a different approach based on the Hidden Local
Symmetry model~\cite{Benayoun:2011mm}. For a concise review of the
muon $g-2$ status and future see Graziano Venanzoni's contribution to
these Proceedings~\cite{VenanzoniLC11}.

The muon is particularly interesting because possible contributions
from unknown heavier states yield contributions\\[-6mm] 
\ba
a_\mu^\mathrm{NP}=\cC\,{m_\mu^2}/{M_\mathrm{NP}^2}
\ea
where naturally $\cC=O(\alpha/\pi)$ ($\sim$ lowest order $\amu^{\rm
SM}$).  Typical New Physics scales required to satisfy $ \Delta
a_\mu^{\rm NP}=\delta a_\mu$ are
$M_\mathrm{NP}=2.0^{+0.4}_{-0.3}~\tv$, $100^{+21}_{-13}~\gv$ and $5^{+1}_{-1}~\gv$
for $\cC$ = 1, $\alpha/\pi$ and $(\alpha/\pi)^2$, respectively.

Different extensions of the SM yield very different
effects in \mbo{a_\mu} such that \mbo{a_\mu} is a very good monitor to
look for physics beyond the SM. It is not so easy to get substantial
effects with obvious new physics possibilities:
in view that the $\tau$ yields $\amu(\tau) \simeq 42 \power{-11}$ only,
and bounds like $m_L> 100$~GeV for a heavy lepton or $m_{b'}\,
\gapprox\, 200$ GeV for a heavy
quark show that sequential fermions (4th family) would not be able to give
a substantial effect. Similarly, possible $Z'$, $W'$ or leptoquarks,
which have to satisfy bounds like $M_{Z',W'}> 600-800$~GeV,
depending  on the GUT scenario yield tiny effects only. They can be
estimated by rescaling the weak
SM contribution with $(M_W/M_{W'})^2\sim 0.01$, which gives 1\% of $19.5 \power{-10}$, 
too small to be of relevance. More examples have been reviewed in~\cite{Jegerlehner:2009ry}. 

Before the recent results from the LHC, constraints on the mass spectrum
from LEP and the Tevatron already excluded sufficiently light new
states which could produce a 3-4 $\sigma$ effect, unless the coupling
is unusually strong, with the risk that perturbative arguments fail to
be reliable.

\begin{figure}
\centering
\includegraphics[height=3cm]{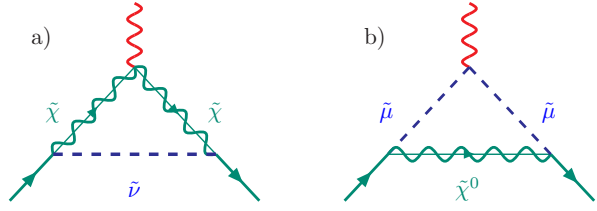}

\caption{Leading SUSY contributions to
$g-2$ in supersymmetric extension of the SM. For subleading
corrections see~\cite{Heinemeyer:2004yq}.
}
\label{fig:SUSYgraphs}
\vspace*{-1cm}
\end{figure}

The most promising New Physics scenarios are provided by SUSY
extensions of the SM because in these models the muon Yukawa coupling
is enhanced by \mbo{\tan \beta =\varv_1/\varv_2} which naturally may be
expected to be as large as the top to bottom quark mass ratio (assuming
$y_t=y_b$) \mbo{\tan \beta =\varv_1/\varv_2=m_t/m_b \sim 40}. Such enhanced
supersymmetric contributions to $a_\mu$ stem from sneutrino--chargino
and smuon--neutralino loops as shown in Fig.~\ref{fig:SUSYgraphs}.
The renormalization group improved 1-loop MSSM result is given by \\[-6mm]
\ba
a_\mu^{\rm SUSY}\!\!
 \simeq   \frac{{\mathrm{sign}(\mu M_2)}\,\alpha(M_Z)}{
8\pi\sin^2\Theta_W}\,}\,\frac{\left(5+\tan^2 \Theta_W \right)}{6}{
\frac{m_\mu^2}{M_{\rm SUSY}^2}\:}{ \tan\beta}{ \: \left( 1-\frac{4\alpha}{
\pi}\ln \frac{M_{\rm SUSY}}{ m_\mu}\right)
\ea
with $M_{\rm SUSY}$ a typical SUSY loop mass and the sign is
determined by the Higgsino mass term $\mu$. Obviously, the 3-4
$\sigma$ deviation in muon $g-2$ (if real) requires
$\mathrm{sign}(\mu)$ positive and $\tan \beta$ preferably large. These
are clear cut constraints which cannot be obtained easily based on LHC
data alone. In GUT constrained models where neutralino masses are
constrained by limits on the colored sector from the LHC, typically
now \MSUSY $>$ 500 GeV. If we assume $\delta a_\mu=\Delta
a_\mu^{\mathrm{SUSY}}$ we find $\tan \beta \simeq M^2_{\mathrm{SUSY}}/
(65.5~\gv)^2$, which for
\MSUSY $\simeq$ 500 GeV requires \tanbeta $\simeq 58$ (see
Fig.~\ref{fig:tanbetaSUSY}), which is in
conflict with a Higgs near 125 GeV as we will see below. 

\begin{figure}
\centering
\includegraphics[height=6cm]{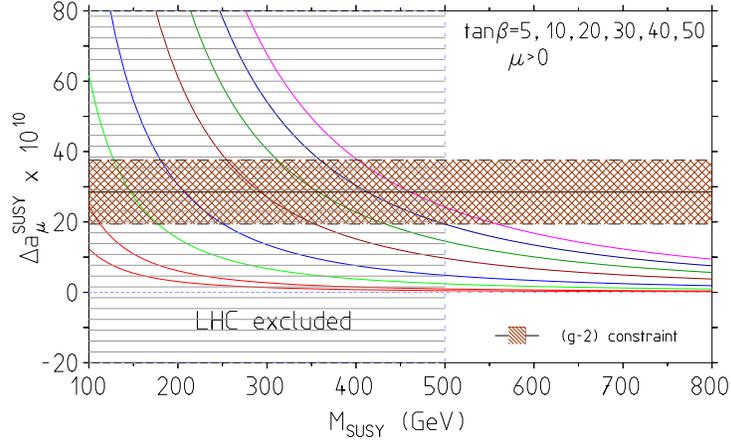}
\caption{Constraint on large \tanbeta SUSY contributions as a
function of $M_{\mathrm{SUSY}}$. The horizontal band shows 
$\Delta a^{\mathrm{NP}}_\mu=\delta
a_\mu$. The region left of \MSUSY $\sim$ 500 GeV is excluded by LHC searches.
If $m_h \sim 125\pm 1.5~\gv$ actually $M_{\mathrm{SUSY}} >
800~\gv$ depending on details of the stop sector
($\{\tilde{t}_1,\tilde{t}_2\}$ mixing and mass splitting) and weakly on \tanbeta.} 
\label{fig:tanbetaSUSY} 
\vspace*{-1cm}
\end{figure}

\section{High energy precision physics: LEP, B-physics}
Here we are looking at SM precision observables like  $G_F$ (muon lifetime), $Z$ observables $M_Z$, $\Gamma_Z$, $g_V$, $g_A$,
$\sin^2 \Theta_{\mathrm{eff}}$ (LEP1/SLD) $W$ and $top$ observables
$M_W$, $\Gamma_W$, $m_t$ and $\Gamma_t$ (LEP2/Tevatron). An important
observable is the $W$ mass given by 
\be
M_W^2\,\left(1-\frac{M_W^2}{M_Z^2}\right)=\frac{\pi \alpha}{\wz
G_F}\,\left(1+\Delta r\right)
\ee
where $\Delta r = f(\alpha,G_F,M_Z,m_t,\cdots)$ represents the
radiative correction to the tree level mass-coupling relation, which
depends on the independent parameters of the theory. They differ from
the SM by additional contributions in extensions of the SM and thus allow to constrain
the parameter space of the extended model. In SUSY models $M_W$ is
sensitive to the top/stop sector parameters and actually $M_W$ is essentially
the only observable which slightly improves in MSSM fits (see Fig.~25 of~\cite{Heinemeyer:2006px})
while 
\be
\sin^2 \Theta_{\mathrm{eff}}=\frac14\,\left(1-\Repa \frac{\varv_{\mathrm{eff}}}{a_{\mathrm{eff}}} \right)
\ee 
remains unaffected~\cite{Buchmueller:2007zk} (see
Figs.~14 and 15 of ~\cite{Heinemeyer:2007bw} and Fig.~1 of
~\cite{Buchmueller:2009ng} and Fig.~4 of
\cite{Buchmueller:2007zk}). The global fit of LEP
data~\cite{lepewwg} does not improve when going from the SM to the MSSM, i.e. SUSY
effects are strongly constrained here. MSSM results merge into SM
results for larger SUSY masses, as decoupling is at work.

Data on the penguin loop induced  $B \to X_s \gamma$ transition 
(see Fig.~\ref{fig:bsg})
\begin{figure}
\centering
\includegraphics[height=4cm]{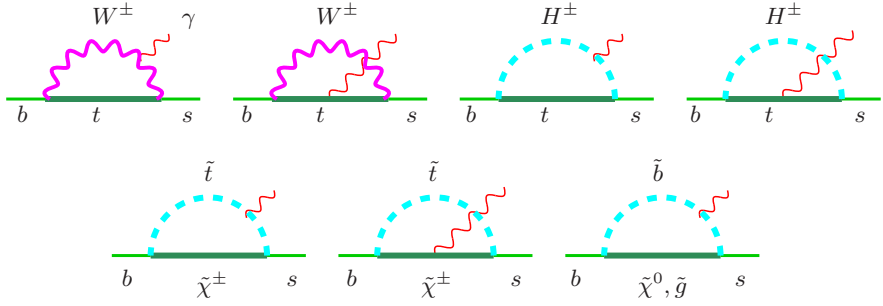}
\caption{Leading graphs in $b \to s \gamma$. SM, 2HDM and SUSY
specific contributions.}
\label{fig:bsg} 
\end{figure}
yields another strong constraint on deviations from the SM~\cite{Bertolini:1990if}.
Indeed, the 
SM prediction~\cite{bsgth}
\mbo{\cB (b \to s\gamma)_{\mathrm{NNLL}} = (3.15 \pm 0.23 )\power{-4}}
is consistent within 1.2 $\sigma$ with the experimental result~\cite{HFAG}
\mbo{\cB (b \to s\gamma) = (3.55 \pm 0.24\pm0.09)\power{-4}}.
It implies that SUSY requires heavier \mbo{m_{1/2}} and/or \mbo{m_0} in order not to
spoil the good agreement.

The very rare box loop induced decay $B_s \to \mu^+\mu^-$ 
(see Fig.~\ref{fig:bsmumu})
\begin{figure}
\centering
\includegraphics[height=6cm]{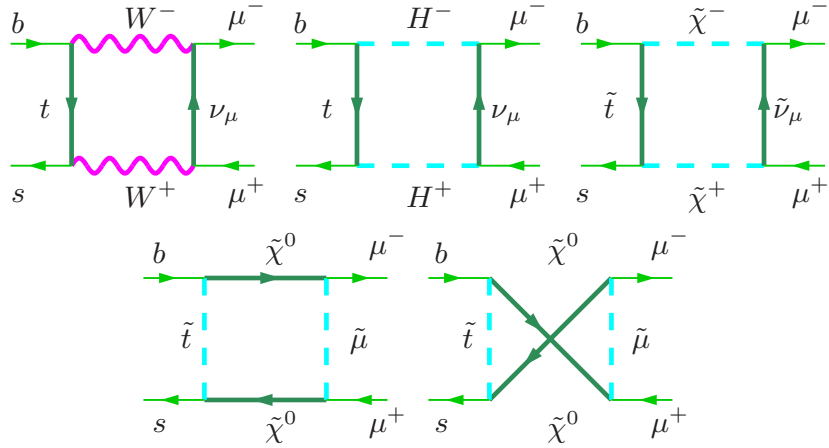}
\caption{Leading graphs in $B_s \to \mu^+\mu^-$. SM, 2HDM and SUSY
specific contributions.}
\label{fig:bsmumu} 
\vspace*{-6mm}
\end{figure}
is very interesting because SUSY contributions
(box contributions with $W$'s replaced by charged Higgses $H^\pm$)
are able to enhance the SM value 
\be\label{SM:pred:BRbmumu}
\branch (\B_s\to \mu^+\mu^-)= (3.1 \pm 1.4) \times 10^{-9}
\ee 
by two orders of magnitude, especially
in scenarios with non-universal Higgs masses (NUHM).
The best bound obtained recently by LHCb~\cite{LHCb:2011ac} is
\be\label{SM:expe:BRbmumu}
\branch (\B_s\to \mu^+\mu^-)< 1.4 \times 10^{-8}\cs
\ee 
and gets closer to the SM value again constraining too large effects
from beyond the SM.

In SUSY+R-parity scenarios dark matter relict density~\cite{Komatsu:2010fb}
\mbo{\Omega_{\mathrm{CDM}} h^2 = 0.1126 \pm 0.0081} represent a tough constraint
for the relic density of neutralinos produced in the early universe.
A DM neutralino is a WIMP DM candidate. The density predicted is~\cite{MicroMegas}
\be
\Omega h^2 \sim \frac{0.1 \,
\mathrm{pb}}{\braket{\sigma\varv}}\sim 0.1 \,\left(\frac{M_{\mathrm{WIMP}}}{100\gv}\right)^2\cs
\ee
where $\braket{\sigma\varv}$ is the relativistic thermally averaged
annihilation cross-section. In most scenarios the dominating
neutralino annihilation process is $\chi+\chi \to A \to b\bar{b}$ and
the observed relict density requires the cross section to be tuned to
$\braket{\sigma\varv}\sim
2\power{-26}~\mathrm{cm}^3/\mathrm{s}$~\footnote{The cross section is
of the form
$$\braket{\sigma\varv}\propto \tan^2 \beta\,
\frac{m_b^2}{M_Z^2}\,\frac{M_\chi^4}{(4M_\chi^2-M_A^2)^2+M_A^2\Gamma_A^2}$$
and has to be adjusted to $M_\chi \approx 1.8\, M_A \mathrm{ \ to \ }2.2\,
M_A$. On resonance the cross section would be
too big, too far off resonance too small~\cite{Beskidt:2012bh}.}. Note that except from
$\Omega_{\mathrm{CDM}}$ all observables prefer heavier SUSY masses
such that effects are small by decoupling. The muon $g-2$ in contrast
requires moderately light SUSY masses and in the pre-LHC era fitted
rather well with expectations from SUSY (see e.g. Fig.~2
of~\cite{Olive:2008uf} and~\cite{Ellis:2009ai}).

\section{Implications of LHC data}
Direct LHC search limits have been taken into account above in some
cases.  LHC events most directly test the colored sector of the
MSSM. In models like the CMSSM and NUHM, constrained by coupling
unification at the GUT scale, the colored sector parameters get
closely related to the uncolored sector. Consequently one obtains
model dependent constraints on physics controlled via standard
precision tests. Typically, in constrained models LHC data have a
strong influence on a large part of SUSY parameter
space~\cite{Buchmueller:2011sw}. The impact is very well illustrated
e.g. in Figs.~1 of~\cite{Olive:2012it,Ellis:2012aa}.
 
A particular role is played by the mass of the light Higgs. At tree level
in the MSSM $m_{h} \leq M_Z$. This bound receives large
radiative corrections from the $t/\tilde{t}$ sector, which changes the upper 
bound to~\cite{HaHe90} \\[-6mm]
\ba 
m^2_{h} \sim M_Z^2\,\cos^2 2\beta + \frac{3\sqrt{2}G_\mu\,m_t^4}{2\pi^2\,\sin^2 \beta}\,\ln
\left(\frac{m_{\tilde{t}_1}\,m_{\tilde{t}_2}}{m_t^2} \right) + \cdots
\ea
which in any case is well below 200 GeV. A given value of $m_{h}$
fixes the value of $m_{1/2}$ represented by
$\{m_{\tilde{t}_1},m_{\tilde{t}_2}\}$. Global frequentist fits to the
CMSSM and NUHM1 scenarios predict $m_{h}\sim 119~\gv$ in fits
incorporating the $(g-2)_\mu$ constraint and $\sim 126~\gv$ without
it. If $m_{h}\simeq 125~\gv$ as suggested by most recent LHC Higgs
searches~\cite{ATLAS,CMS} $m_{1/2}$ would be fixed around 800 to 950
GeV. Together with the narrow bound from the cosmic relict density in
the CMSSM one would also fix $m_0$ at a relatively low value depending
sensitively on $\tan \beta$, however.

As we see the present LHC data have a quite dramatic impact on SUSY
scenarios.  The main outcome is that in constrained models like
CMSSM, NUHM1, mSUGRA or NUHM2 all allowed parameter points with $m_h \sim 125~\gv$
are inconsistent with the observed $(g-2)_\mu$~\cite{Baer:2011ab,Ellis:2012aa} !

\section{Comments and Outlook}
SUSY is the natural mechanism to tame the cosmological constant
problem as well as the Higgs hierarchy problem of the SM. However, to
make a SUSY extension of the SM not to spoil phenomenologically
established minimal flavor patterns of the SM one has to supplement it
by assuming R-parity as an extra symmetry. Most of the popular MSSM scenarios
assume in addition GUT to be at work which heavily constrains the
parameter ambiguities in the possible soft SUSY breakings. One should be
aware of the fact that SUSY and GUT are uncorrelated symmetry
concepts, GUT assumptions almost always made in SUSY extensions of the
SM may not be realistic. Unlike chiral symmetry, gauge symmetry and
super symmetry gauge unification is not imposed to protect light
states since the GUT scale is only two or three orders of magnitude below
$M_{\mathrm{Planck}}$.

Another comment concerns the nature of dark matter. DM is quite
commonly assumed to consist of one or several species of elementary
particle. In the SUSY+R-parity framework the LSP is an elementary
particle. Here we should keep in mind that normal matter in the
universe is dominated by nucleons, i.e., the normal matter density is
99\% frozen energy and the light fermion masses generated in
electroweak symmetry breaking represent a minor correction only. What
if dark matter is not the result of the existence of a stable heavy
elementary particle, but again mostly a form of frozen energy? One
could think of unflavored SU(4) confined states. Such dark matter
would be bosonic with no new fermionic matter which would form DM
stars.  Stability of such matter would be natural similar to B
conservation for normal baryonic matter. In this context direct DM
searches~\cite{Aprile:2011hi} are extremely important and progress in
this field will bring more light into the DM puzzle.

Before the LHC was in operation one was expecting that SUSY improves
agreement with experiment for observables like $a_\mu$ and marginally
for $M_W$.  After the first LHC results the expectations have
changed. The situation looks somewhat disturbing. If a Higgs of mass
near 125 GeV was confirmed one would have a strong point for SUSY at
work\footnote{There is in fact a very different scenario which
predicts a Higgs of mass in this region: Schlereth's
model~\cite{Schlereth:1990iv} of composite weak bosons, exhibiting the
weak gauge bosons as vector mesons and the Higgs as a weak version of
the $\eta'$ predicts a Higgs boson of mass somewhat above the ones of
the gauge bosons~\cite{Galli:1994my}}. But the $(g-2)_\mu$ deviation
requires unexpectedly large \tanbeta now. Trivially, $\tan \beta >
m_t/m_b$ requires that the top-bottom Yukawa couplings must exhibit an
inverse hierarchy $y_b > y_t$ relative to the masses, which to me
looks quite unnatural.

I start to worry about the muon anomaly result in the sense that it is
not 100\% clear to me whether experiments measure what theoreticians
calculate, namely $a_\mu=F_2(0)$?  The fact that perfect charged
one-particle states do not exist, due to the infrared problem of QED,
could affect the measurement at the level of precision we are
dealing with. To my knowledge, in deriving the equations of motion
of the muon in the external field the radiation field is
neglected. The possible problem has been investigated at leading order
in~\cite{Steinmann02} some time ago, but no higher order results have
been worked out so far. At the given level of precision this is an issue
which should be investigated more carefully.

Within the next 5 years a new muon $g-2$ experiment will go into
operation at Fermilab (E989)~\cite{FermilabE989}. It will be an
upgraded Brookhaven experiment (E821) working with ultra-relativistic
magic energy muons. An alternative project is being designed at
J-PARC which will work with ultra-cold muons~\cite{JPARCg-2}. The
experiment will have very different systematics and therefore will
provide a very important crosscheck of the storage ring
experiments. The accuracy attempted is \mbo{\delta
a_\mu=16\power{-11}}. Provided the deviation (\ref{deviation}) is real
and the central value would not move the $3\sigma$ would turn into a
$9\sigma$ deviation, provided theory is able to reduce theoretical
uncertainties accordingly.  If SUSY or 2HDM would be at work this experiment
would provide invaluable information about the sign of the parameter
$\mu$ and pin down $\tan \beta$ like no other experiment~\cite{Hertzog:2007hz}.

One has to be aware that much of the tension in the interpretation of
the data we are confronted with may be a result of too special model
assumptions used in analyzing the data.  First LHC data primarily
constrain the colored sector. In those SUSY models which do
\underline{not assume}
a strong correlation between the colored and the uncolored sector a
future ILC(1000) would play a prominent role in disentangling the
true structure beyond the SM. For much more detailed discussions
I refer to~\cite{others} beside the articles quoted earlier.

\section*{Acknowledgments}
I am grateful to the organizers for the kind invitation, for the support and the
kind hospitality at the ECT* Trento. I thank Rainer Sommer for
useful comments and for carefully reading the manuscript.

 \end{document}